\documentclass[aps,prd,twocolumn,groupedaddress,nofootinbib]{revtex4}  
\usepackage{graphicx}  
\usepackage{dcolumn}   
\usepackage{bm}        
\usepackage{amssymb,amsfonts,amsmath,physics}   
\usepackage{slashed}

\usepackage{tikz-feynman}
\tikzfeynmanset{compat=1.1.0}
\usetikzlibrary{arrows}
\tikzset{
    vertex/.style={circle,draw, minimum size=1.5em},
    edge/.style={->,> = latex'}
}

\usepackage[bookmarks, breaklinks, colorlinks,urlcolor=blue, citecolor=red, 
linkcolor=blue]{hyperref}

\usepackage[normalem]{ulem}

\def\a {\alpha}
\def\S {\Sigma}
\def\D {\Delta}

\def\ve {\varepsilon}
\def\n {\nonumber}

\def\r {\rightarrow}
\def\b {\beta}
\def\g {\gamma}

\def\u {\upsilon}
\def\l {\lambda}

\def\bar {\overline}

\def\lr {\leftrightarrow}

\def\abs#1{\left| #1 \right|}

\def\uo {\upsilon_1}
\def\ut {\upsilon_2}
\def\lfi {\lambda_5}

\def\mtt {m_{22}}

\def \ms {M_{\S_i}}
\def \gtr {\g_A/n_\S^{eq}H}
\def \idtr {\g_D/n_l^{eq}H}
\def \mt {\tilde{m}}
\def \pt {\Phi_2}

\newcommand{\bmt}{\begin{pmatrix}}
\newcommand{\emt}{\end{pmatrix}}
\newcommand{\ba}{\begin{array}{c}}
\newcommand{\ea}{\end{array}}
\newcommand{\be}{\begin{equation}}
\newcommand{\ee}{\end{equation}}
\newcommand{\bea}{\begin{eqnarray}}
\newcommand{\eea}{\end{eqnarray}}

\newcommand{\bi}{\begin{itemize}}
\newcommand{\ei}{\end{itemize}}

\newcommand{\baz}{\begin{array}{cc}}
\newcommand{\mathsym}[1]{{}}

\newcommand{\bt}{\begin{tabular}}
\newcommand{\et}{\end{tabular}}

\newcommand{\benu}{\begin{enumerate}}
\newcommand{\eenu}{\end{enumerate}}

\begin{document}

\title{Lowering the scale of fermion triplet leptogenesis with two Higgs doublets}

\author{Drona Vatsyayan}
\email[E-mail: ]{drona.vatsyayan@ific.uv.es}
\affiliation{Departamento de F\'isica Te\'orica, Universidad de Valencia and IFIC, Universidad de Valencia-CSIC,
C/ Catedr\'atico Jos\'e Beltr\'an, 2 | E-46980 Paterna, Spain}

\author{Srubabati Goswami}
\email[E-mail: ]{sruba@prl.res.in}
\affiliation{Theoretical Physics Division, Physical Research Laboratory, Ahmedabad-380009, India}


\begin{abstract} 

In this paper we consider  the possibility of generating   the observed baryon asymmetry of the  Universe via leptogenesis  in the context of a triplet fermion-mediated type-III seesaw  model of neutrino mass. With a  hierarchical spectrum of the additional fermions, the lower bound on the lightest triplet mass is $\sim 10^{10} {\rm~GeV}$ for successful leptogenesis, a couple of orders higher than that of type-I case. We investigate the possibility of lowering this bound in the framework of two-Higgs-doublet models. We find that the bounds can be lowered down to $10^7$ GeV for a hierarchical spectrum. If we include the flavor effects, then a further lowering by one order of magnitude is possible. We also discuss if such lowering can be compatible with the naturalness bounds on the triplet mass. 
\end{abstract}

\maketitle


\section{Introduction}
\label{sec:intro}

The Standard Model (SM) is immensely successful in explaining majority of the  properties of the fundamental particles  and their interactions.  The SM was further bolstered by discovery of the Higgs boson by the ATLAS and CMS experiments of the Large Hadron Collider (LHC)~\cite{ATLAS:2012yve, CMS:2012qbp}. However, there are  some experimental results that cannot be accounted for in the framework of the SM. Two of the most relevant of these for our purposes are as follows: (i) The SM fails to explain the origin of the baryon asymmetry of the Universe, $Y_{\Delta B}=(8.75 \pm 0.23) \times 10^{-11}$, determined from cosmic microwave background radiation and relative abundances of light elements~\cite{Planck:2018vyg}. (ii) On the other hand, several neutrino oscillation experiments have shown that neutrinos have tiny but nonzero masses, unlike what is predicted by the SM. One of the most appealing scenarios to explain  the tiny neutrino masses needed to explain the results from oscillation experiments is the seesaw mechanism which is generated by the dimension-5 Weinberg operator~\cite{Weinberg:1979sa}. The ultraviolet  completion of this can be accomplished in three ways  giving rise to three different versions of the model.  These are  achieved by  introduction of heavy right-handed neutrinos (type I), scalar triplets (type II), or fermion triplets (type III). Since the Weinberg operator violates lepton number by 2 units, seesaw implies that neutrinos are Majorana particles.

In the type-I seesaw model, the ou-of-equilibrium decays  of the heavy Majorana  neutrinos can give rise to a lepton asymmetry (leptogenesis) which can be subsequently reprocessed into baryon asymmetry via sphaleron processes~\cite{Fukugita:1986hr}.   For thermal leptogenesis with a hierarchical spectrum of heavy neutrinos,  generation of required CP asymmetry  puts  a lower bound on the mass of the lightest right-handed neutrino. This is the famous Davidson-Ibarra bound  which gives $M_{\rm lightest} > 10^8 - 10^9  $ GeV~\cite{Davidson:2002qv}. For the case of the type-III seesaw model,  in which the lepton asymmetry is generated from the CP-violating decays of fermion triplets, the lower bound on the CP asymmetry combined with the fact that the $SU(2)$ triplet fermions have additional gauge interactions increases the above bound to $M_{\rm lightest} > 3 \times 10^{10} $ GeV~\cite{Hambye:2003rt}. 

These bounds are in tension with the constraints coming from naturalness, implying that the presence of these heavy particles can lead to large quadratic corrections to the Higgs mass, enhancing the hierarchy problem. If we assume naturalness to exist, it puts an upper limit on the mass scale of the seesaw mediators. The naturalness constraints in the context of the minimal type-III seesaw model were studied in Ref.~\cite{Goswami:2018jar}. For instance, assuming that the quadratic corrections to Higgs mass are $< 1 {\rm~TeV^2}$, they obtained $M_{\rm lightest} \leq 3 \times 10^5 - 1.84 \times 10^7 {\rm~GeV}$, depending on the parameters of the model. Current searches for such fermion triplets at the ATLAS~\cite{ATLAS:2020wop} and CMS~\cite{CMS:2019lwf} detectors at the 13 TeV LHC have set the lower limits on their mass at $790$ and $880 {\rm~GeV}$ at $95\%$ C.L. respectively.

In literature, attempts have been made to reduce the bound on the lightest seesaw mediator mass by adding extra scalar fields~\cite{Goswami:2021eqy, Clarke:2015hta, Hugle:2018qbw, Alanne:2018brf}. In particular, Ref.~\cite{Clarke:2015hta} considered a two-Higgs-doublet model (2HDM) in the context of the type-I seesaw and they showed that the scale of leptogenesis can be lowered and made consistent with the naturalness bound if the second Higgs doublet gets a very small vacuum  expectation value (VEV). In this paper, we investigate if a similar phenomenon can happen for the type-III seesaw model. In particular, we examine the role of gauge scatterings, and an additional doublet in lowering the leptogenesis scale. In addition, we also include the flavor effects and see the implications of these in lowering the leptogenesis scale further. 

The paper is structured as follows. In Sec.~\ref{sec:model}, we discuss the two-higgs-doublet model with triplet fermions. In Sec.~\ref{sec:lepto}, we briefly discuss type-III leptogenesis in the usual case with just one Higgs doublet and present the numerical results for the 2HDM case in Sec.~\ref{sec:result}. We summarize our findings in Sec.~\ref{sec:con}.

\section{Model}
\label{sec:model}

We extend the Standard Model by three hierarchical fermionic $SU(2)$ triplets $\Sigma_i (i=1,2,3)$ with zero hypercharge $(M_{\S_3} \gg M_{\S_2}\gg M_{\S_1})$ and the scalar sector by an additional Higgs doublet with hypercharge one. Thus, our model contains two $SU(2)$ doublets $\Phi_{i}(i=1,2)$. Since the general 2HDM has tree-level Higgs-mediated flavor-changing neutral currents (FCNCs) (contradicting the experimental bounds on FCNCs), a discrete $Z_2$ symmetry is imposed on the scalar potential as well as to the fermions to avoid such currents~\cite{Accomando:2006ga}. These FCNCs can also be eliminated if the $Z_2$ symmetry is softly broken by demanding that each right-handed fermion couples to only one of the doublets~\cite{Glashow:1976nt, Paschos:1976ay}. Here, we assume that $\Phi_1$ ($\Phi_2$) couples only to $u_R^i$ ($\Sigma_i$).  

The CP-conserving, softly broken $Z_2$ symmetric potential in the 2HDM model can be written as
\begin{align}\label{eq:scapot}
    V(\Phi_1,\Phi_2) &= m_{11}^2 \Phi_1^\dagger \Phi_1 + m_{22}^2 \Phi_2^\dagger \Phi_2 - m_{12}^2 (\Phi_1^\dagger \Phi_2 + \Phi_2^\dagger \Phi_1)\n\\
    &+\frac{\lambda_1}{2}(\Phi_1^\dagger \Phi_1)^2 + \frac{\lambda_2}{2}(\Phi_2^\dagger \Phi_2)^2 \n\\
    &+ \lambda_3 (\Phi_1^\dagger \Phi_1)(\Phi_2^\dagger \Phi_2)+\lambda_4 (\Phi_1^\dagger \Phi_2)(\Phi_2^\dagger \Phi_1)\n \\
     &+\frac{\lambda_5}{2}[(\Phi_1^\dagger \Phi_2)^2+(\Phi_2^\dagger \Phi_1)^2]\,,
\end{align}
where all the parameters can be chosen real without loss of generality. The $m_{12}^2$ term with operator dimension two above softly breaks the $Z_2$ symmetry.\footnote{This implies that the $Z_2$ symmetry is respected at short distances i.e. at distances shorter than the cut-off $1/M$ to all orders in the perturbative series. Thus, in the limit $M^2 \r \infty$ or equivalently at high energies, the $Z_2$ symmetry is restored \cite{Ginzburg:2004vp}.}

After electroweak symmetry breaking, the scalar fields can be decomposed as
\begin{align}
    \Phi_i=
    \begin{pmatrix}
    \phi_i^+ \\
    {(\upsilon_i + \rho_i +i \eta_i)}/{\sqrt{2}}
    \end{pmatrix},
\end{align}
where $\upsilon_i$ are the VEVs given by
\begin{align}\label{eq:vev}
\uo &\approx \sqrt{\frac{-2m_{11}^2}{\l_1}}\,,\n\\
\ut &\approx \frac{1}{1+\frac{\uo^2}{2m_{22}^2}\l^{\ast}}\frac{m_{12}^2}{m_{22}^2}\uo\,,
\end{align} 
with $\l^{\ast}=\l_3+\l_4+\l_5$, $m_{11}^2 < 0$, and $\uo^2+\ut^2=\u^2 \approx (246 {\rm GeV})^2$. 

From Eq.~(\ref{eq:vev}), we can see that the vev $\ut$ can be very small if $m_{12}^2/m_{22}^2 \ll 1$. This is natural in the t'Hooft sense since, as $m_{12}^2 \rightarrow 0$, a $U(1)$ or $Z_2$ symmetry is reinstated for $\lambda_5 =0$ or $\lambda_5 \neq 0$ respectively. Further, the minimization conditions of the scalar potential lead to an upper bound on $m_{22}^2$~\cite{Clarke:2015hta}
\begin{equation}\label{eq:cons}
m_{22}^2 \lesssim \frac{1}{2} m_h^2 \tan^2 \beta
\end{equation}
in the limit $m_{22}^2 \gg \l_2 \ut^2$, provided that $m_{11}^2$ remains negative. Here, $m_h = 125 {\rm~GeV}$ is the Higgs mass and $\tan \beta= \uo/\ut$.

The part of the Lagrangian concerning neutrino masses and leptogenesis is given by
\begin{eqnarray}\label{eq:lag}
    \mathcal{L}_{\S}=& {\rm Tr}[\bar{\S_i}i \slashed{D}\S_i] -{\tilde{\Phi}}_2^\dagger \bar{\S_i} \sqrt{2}Y_{\a i} L_\a \n \\
    &-\frac{1}{2}{\rm Tr}[\bar{\S_i}M_{\S_i}\S_i^c] + \text{H.c.}\,,
\end{eqnarray}
where $L_\a=(\nu,l)^T~ (\a=e,\mu,\tau)$ and $\Phi_2=(\phi_2^+,\phi_2^0)^T$ are the lepton and Higgs doublet, respectively, with $\tilde{\Phi} \equiv i \sigma_2 \Phi^\ast$ and $D_\mu \equiv \partial_\mu -ig \tau^A W_\mu^A/2$. The triplets can be represented as
\begin{align}
\S =
\bmt
\S^0/\sqrt{2} &  \S^+ \\
\S^- & -\S^0/\sqrt{2}
\emt\,,
\end{align}
with $\S^{\pm}=(\S^1 \mp i\S^2)/\sqrt{2}$. 

For $m_{12}^2 > 0$ and $\lfi=0$, the $U(1)$ symmetry is softly broken, and one can write the neutrino mass matrix in terms of the usual type-III seesaw formula with $\u$ replaced by $\ut$:
\begin{align}
\mathcal{M}_\nu^{\rm tree} = -\frac{\ut^2}{2}{\bm Y}^T\frac{1}{\bm M_\S}{\bm Y}\,.
\end{align} 
The model also has the possibility of generating neutrino masses radiatively if $m_{12}=0,\lfi \neq 0$. In this case, $\ut =0$ and the $Z_2$ symmetry remains unbroken; therefore, neutrinos cannot have masses at tree level. This case corresponds to the type-III scotogenic Model and the neutrino masses are given by~\cite{Ma:2008cu}
\begin{align}
\mathcal{M}_{\a\b}^{\rm loop}=\frac{Y_{\a i}Y_{\b i}\,\lfi \u^2}{32\pi^2\, \ms}\bigg[\frac{r}{1-r}\bigg( \frac{1}{1-r}\bigg)\ln r \bigg]\,,
\end{align}
where $r=\ms^2/M_\phi^2$ and $M_\phi^2=m_{22}^2 +(\l_3+\l_4)\u^2$.

For further analysis, we consider $\lfi=0$, so that neutrino masses are generated only at the tree level. Leptogenesis in the context of a hybrid model of scotogenic type-I and -III seesaw has been discussed in Ref.~\cite{Suematsu:2019kst}.

\section{Leptogenesis}
\label{sec:lepto}

The mass term for the Majorana triplets in Eq.~(\ref{eq:lag}) breaks the lepton number, and in general, the Yukawa coupling matrix ${\bm Y}$ is complex and contains new sources of CP violation; therefore, if the decay of the fermion triplet $\S \r l \Phi_2$ is out of equilibrium, then all the Sakharov conditions~\cite{Sakharov:1967dj} are satisfied for the dynamic generation of a lepton asymmetry. This is similar to type-I leptogenesis, however, qualitative differences arise due to the coupling of fermion triplets to SM gauge bosons, as we discuss later.

The lepton asymmetry produced by the decay of a heavy triplet can be written as~\cite{Hambye:2012fh}
\begin{align}
Y_L \equiv \ve_\S \eta Y_\S({T\gg M_\S}) = \frac{135 \zeta(3)}{4\pi^4 g_\ast}\ve_\S \eta \,,
\end{align}
where $\ve_\S$ is the average lepton asymmetry $\Delta L$ produced per decay of $\S$ (CP asymmetry parameter), $Y_a \equiv n_a/s$ is the number density of species $a$ per unit entropy, $n_\S$ is the total number of triplets and $g_\ast$ is the number of active degrees of freedom at $T=M_1$. The efficiency factor, $0<\eta \leq 1$ incorporates the effects of washouts and scatterings and is determined from the numerical solution of the Boltzmann equations for the lepton asymmetry and triplet population. The baryon asymmetry produced from this lepton asymmetry is
\begin{align}\label{eq:yb}
Y_B = 3\, C\, Y_{B-L} = -0.0041 \ve_\S \eta\,,
\end{align}
where $C=-8/23$ is the sphaleron conversion factor for three flavors and two Higgs doublets~\cite{Harvey:1990qw}, and the factor of 3 comes from the three $SU(2)$ degrees of freedom of the triplets. 

\subsection{CP asymmetry}
\label{sec:cpasym}


For a hierarchical spectrum of heavy seesaw states $(M_{\S_1}< M_{\S_2}< M_{\S_3})$, only the decays of the lightest state are important for leptogenesis as the asymmetry generated by the heavier states can be washed out by $\Sigma_1$ -mediated interactions until they drop out of equilibrium. Considering the decay of the lightest triplet $\S_1$, CP asymmetry can be calculated as~\cite{Hambye:2003rt}
\begin{align}\label{eq:cpasymm}
\ve_{\S_1}=\sum_{j=2,3}\frac{3}{2}\frac{M_{\S_1}}{M_{\S_j}}\frac{\Gamma_j}{M_{\S_j}}I_j \frac{V_j-2 S_j}{3}\,,
\end{align}
where 
\begin{align}
I_j=\frac{{\rm Im}[(Y^\dagger Y)_{1j}^2]}{\abs{Y^\dagger Y}_{11}\abs{Y^\dagger Y}_{jj}}\,,
\end{align}
and $S_j$ and  $V_j$ are the loop factors associated with the self-energy and vertex corrections, respectively. $\Gamma_j$ corresponds to the decay width of triplet components, which is similar to that of a right-handed neutrino and can be expressed as
\begin{align}
\Gamma_{j}&=\frac{1}{8\pi}M_{\S_j} (Y^\dagger Y)_{jj}(1-y_j)^2\,,
\end{align}  
where the factor $y_j=\mtt^2/M_{\S_j}^2$ takes the mass of the second Higgs doublet into account. 
\begin{figure}[!htb]
\centering
\includegraphics[scale=0.4]{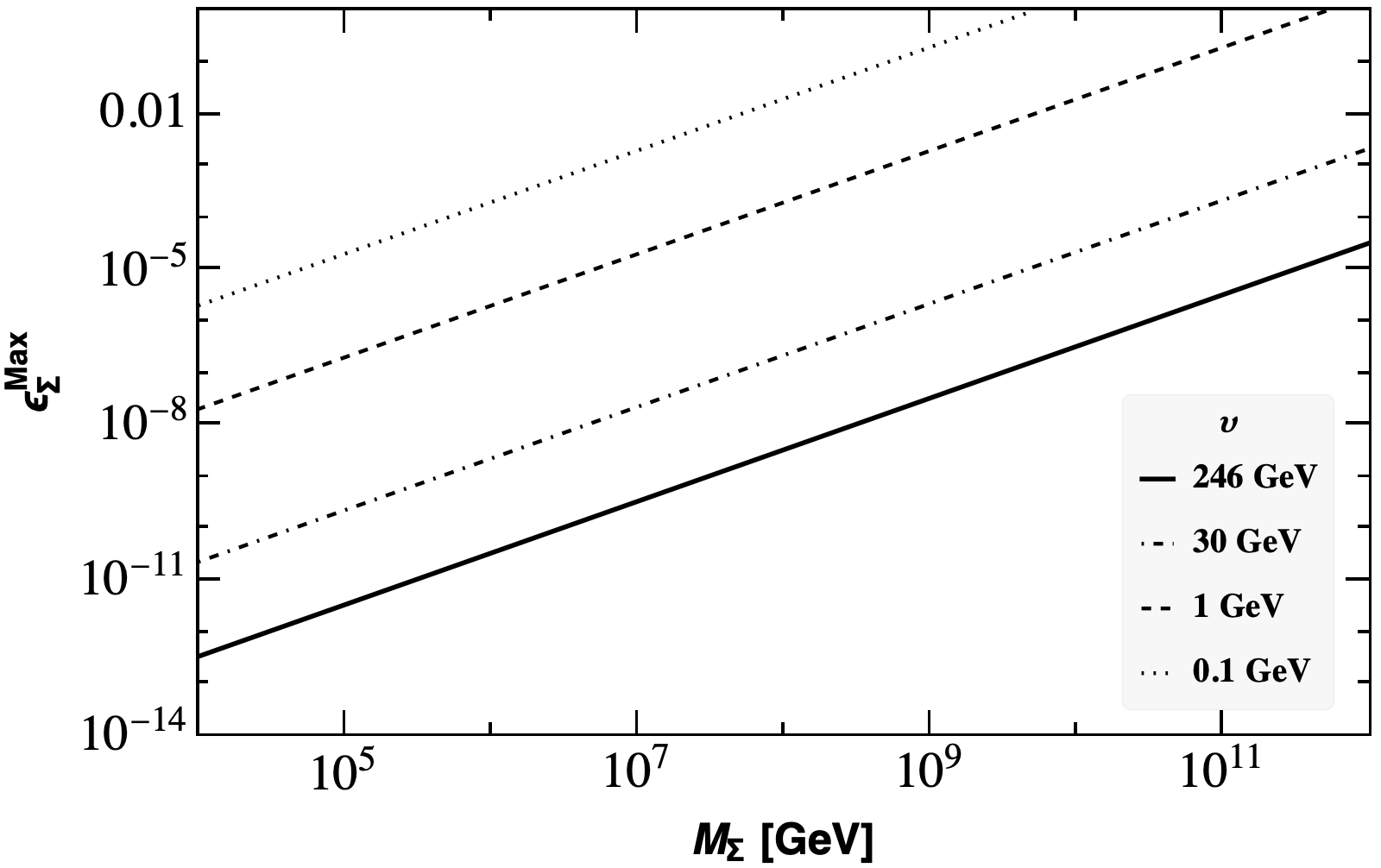}
\caption{The maximum bound on CP asymmetry as a function of triplet mass $M_\S$ for different values of $\ut$. The solid black line represents the standard case with one Higgs doublet and $\u = 246 {\rm GeV}$}
\label{fig:DI}
\end{figure}
In the type-I seesaw, there is an upper bound on CP asymmetry~\cite{Davidson:2002qv} 
\begin{equation}
\ve_N \leq \frac{3}{8\pi}\frac{M_N}{\upsilon^2}(m_{\nu_3} - m_{\nu_1})\,,
\end{equation}
which can be translated into a lower bound on the mass of right-handed neutrinos $M_N \gtrsim 6 \times 10^8 {\rm~GeV}$, assuming maximal efficiency i.e. $\eta=1$. For triplets, the same expression holds with a factor of $1/3$ and $\u$ replaced by $\ut$ in our model and we get
\begin{align}\label{eq:DI}
\ve_\S \leq \frac{1}{8\pi}\frac{M_\S}{\ut^2}(m_{\nu_3} - m_{\nu_1})\,.
\end{align}
Note the inverse dependence on $\ut$, which allows us to relax the bounds on maximum allowed CP asymmetry for smaller values of VEVs as shown in Fig.~\ref{fig:DI}. However, unlike the case of type-I seesaw, we cannot translate this into a lower bound on the triplet mass as gauge scatterings can suppress the efficiency considerably and we cannot assume $\eta=1$. Hence, in order to compute the lower bound for the triplet mass, one must solve the coupled Boltzmann equations for the evolution of number density of triplets and lepton asymmetry.

\subsection{Boltzmann equations}
\label{sec:BE}

Assuming flavor-blind leptogenesis and considering only gauge scatterings mediated by SM gauge bosons, decays (and inverse decays) and $\Delta L =2$ scatterings, the Boltzmann equations can be written as\footnote{Since we are only concerned with the decay of the lightest triplet, we suppress the index 1.}
\begin{align}\label{eq:beqs}
sHz\,\frac{dY_\S}{dz}&=-\bigg(\frac{Y_\S}{Y_\S^{eq}}-1\bigg)\g_D - 2\bigg(\frac{Y_\S^2}{Y_\S^{2eq}}-1\bigg)\g_A\,, \\ \label{eq:beql}
sHz\frac{dY_{B-L}}{dz}&=-\g_D \ve_\S \bigg(\frac{Y_\S}{Y_\S^{eq}}-1\bigg)-\frac{Y_{B-L}}{Y_l^{eq}}\bigg(\frac{\g_D}{2}+2\g_\S^{\rm sub}\bigg)\,,
\end{align}  
where $Y_a = n_a/s$ is the yield, $z \equiv M_\S/T$ is the evolution variable, $H=1.66 \sqrt{g_\ast}\,T^2/M_{\rm Pl}$ is the Hubble expansion rate of the Universe and $s=0.44\,g_\ast T^3$ is the entropy density. The suffix `eq' denotes the equilibrium values and $\g$'s are the reaction densities for various processes: $\g_D$ is the decay/inverse decay reaction rate, $\g_A$ is the gauge scattering reaction rate for the processes $\S\bar{\S}'\lr GG',f\bar{f},H\bar{H}$, and $\g_\S^{\rm sub}$ comes from $\Delta L=2$ scatterings $LL \lr \pt^\ast \pt^\ast,L\pt \lr \bar{L}\pt^\ast$ with the on-shell part subtracted.

The decay and gauge scattering densities are given by~\cite{Hambye:2003rt,AristizabalSierra:2010mv}
\begin{align}
\g_D(z) &= \frac{1}{8\pi^3}\frac{M_\S^5}{\ut^2}\frac{K_1(z)}{z}\tilde{m}\,,\\
\g_A(z) &=\frac{M_\S^4}{64\pi^4}\int_4^\infty dx \frac{\sqrt{x}K_1 (z\sqrt{x})\hat{\sigma}_A(x)}{z}\,,
\end{align}
where $x =s/M_\S^2$, $s$ is the center of mass energy squared and $K_1(z)$ is the modified Bessel function of the first kind. We define
\begin{align}
\tilde{m}=\frac{(Y^\dagger Y)_{11}}{2 M_\S}\ut^2 = 8\pi \Gamma_\S \frac{\ut^2}{M_\S^2}
\end{align}
as the effective neutrino mass and use the following expression for the reduced cross section $\hat{\sigma}_A (x)$~\cite{Hambye:2003rt} 
\begin{align}
\hat{\sigma}_A (x)&=\frac{6g^4}{\pi}\bigg(1+\frac{2}{x}\bigg)r +\frac{2g^4}{\pi}\bigg[ -\bigg(4+\frac{17}{x}\bigg)r \n\\
&+3\bigg(1+\frac{4}{x}-\frac{4}{x^2}\bigg)\log\bigg(\frac{1+r}{1-r}\bigg)\bigg]\,,
\end{align}
with $r=\sqrt{1-4/x}$. Note that since the gauge scatterings do not violate $L$, they affect only the evolution of $\S$ population. In the absence of $\g_A$, the Boltzmann equations take the exact same form as that for right-handed neutrinos, with all the interaction terms multiplied by 3 in Eq.~(\ref{eq:beql}). It can be seen that the decay and gauge scattering rates depend only on $\tilde{m}$ and $M_{\S}$, which we will take as free parameters for our analysis.\footnote{Measuring neutrino masses does not fix $\tilde{m}$ and $M_\S$ and hence they can be treated as free parameters. See Ref.~\cite{Giudice:2003jh}}

\subsection{Gauge versus Yukawa interactions}
\label{sec:gaugeyukawa}

\begin{figure}
\centering
\includegraphics[scale=0.4]{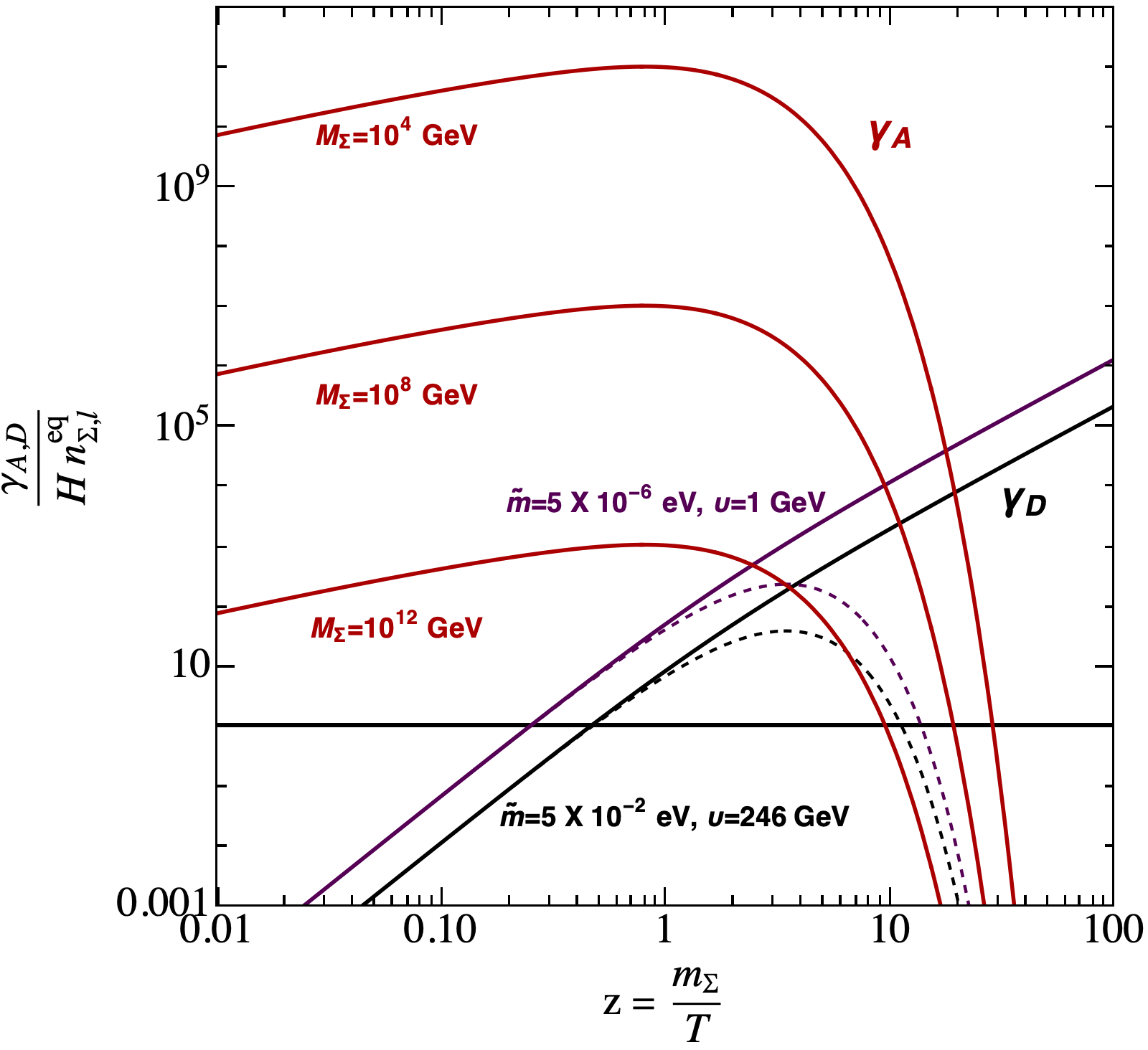}
\caption{Evolution of gauge and Yukawa reaction densities relevant for Boltzmann equations. The decay (inverse decay) thermalization rate $\g_D/H n_{\S,(l)}^{eq}$ denoted by solid(dashed) curves depends only on $\tilde{m}$ for a particular value of vev (black, purple), whereas the gauge thermalization rates $\g_A/H n_\S^{eq}$ denoted by red curves depend only on $M_\S$. The horizontal black line indicates when the reactions go out equilibrium.}
\label{fig:int}
\end{figure}

Now, we discuss the effect of gauge reactions which are a function of $M_\S$ only. At high temperatures $z\ll 1$, the gauge reactions are in equilibrium and tend to thermalize the triplet population; thus, no asymmetry can be generated. However, at low temperatures $z \gg 1$, the gauge reaction rate is doubly Boltzmann suppressed and an asymmetry can be generated when these gauge reactions decouple, which implies
\begin{align}
\frac{\Gamma_A}{H}=\frac{\g_A}{n_\S^{eq}H} \lesssim 1\,,
\end{align}
provided that the inverse decays $l\Phi_2 \r \S$ are also decoupled at that stage. If the inverse decays are still active, the asymmetry will be generated at a later stage once they go out of equilibrium, i.e., $\g_D/n_l^{eq}H \lesssim 1$. To understand the dynamics better, we plot the thermalization rates for various values of $M_\S,\tilde{m}$ and $\ut$ in Fig.~\ref{fig:int}.

As can be seen from Fig.~\ref{fig:int}, the gauge thermalization rate $\Gamma_A/H=\g_A/n_\S^{eq}H$ reaches its maximum value at $z=1$, i.e., $T \sim M_\S$, and the lower the triplet mass, the more the triplet is thermalized. As long as $\gtr>1$, $Y_\S$ tracks its equilibrium distribution $Y_\S^{eq}$, and the gauge interactions dominate over the Yukawa ones for $\g_D <4\g_A$,  i.e. triplets scatter before they decay. The asymmetry is, thus, suppressed by a factor of $\g_D/4\g_A$ in this period until either $\gtr < 1$ or $4\g_A < \g_D$, where the latter condition is used to define $z=z_A$~\cite{Hambye:2012fh}. 

For $z>z_A$, the decays dominate over the gauge interactions and the triplets decay before scattering. Hence, there is no suppression from gauge scatterings, even if they are in equilibrium. The only suppression comes from the number of triplets that survive gauge annihilations at $z=z_A$, which is a Boltzmann-suppressed quantity. For the SM VEV (shown in black in Fig.~\ref{fig:int}), the value of $z_A$ lies around $~4$ for $M_\S=10^{12} {\rm~GeV}$ and $~21$ for $M_\S=10^{4}{\rm~GeV}$. Notice that the larger the value of $z$, the lower the number of triplets left at that point to generate the lepton asymmetry. Therefore, successful low-scale type-III leptogenesis becomes extremely difficult. 

The washout from inverse decays, characterized by the thermalization rate $\idtr$ is depicted by the dashed lines in Fig.~\ref{fig:int}. The inverse decays are suppressed at $z>1$, so, if $\idtr <1$ at $z=z_A$, the inverse decays are already decoupled and their contribution to washout can be neglected, and the major contribution to washout comes from gauge reactions. This usually holds true for smaller values of $\mt$, where $\idtr$ never reaches unity. Larger values of $\mt$ can keep inverse decays in equilibrium at $z \geq z_A$, where their contribution to the washout starts dominating over that of gauge reactions and the efficiency is suppressed mainly by inverse decays.

\begin{figure}[!htb]
\centering
\includegraphics[scale=0.4]{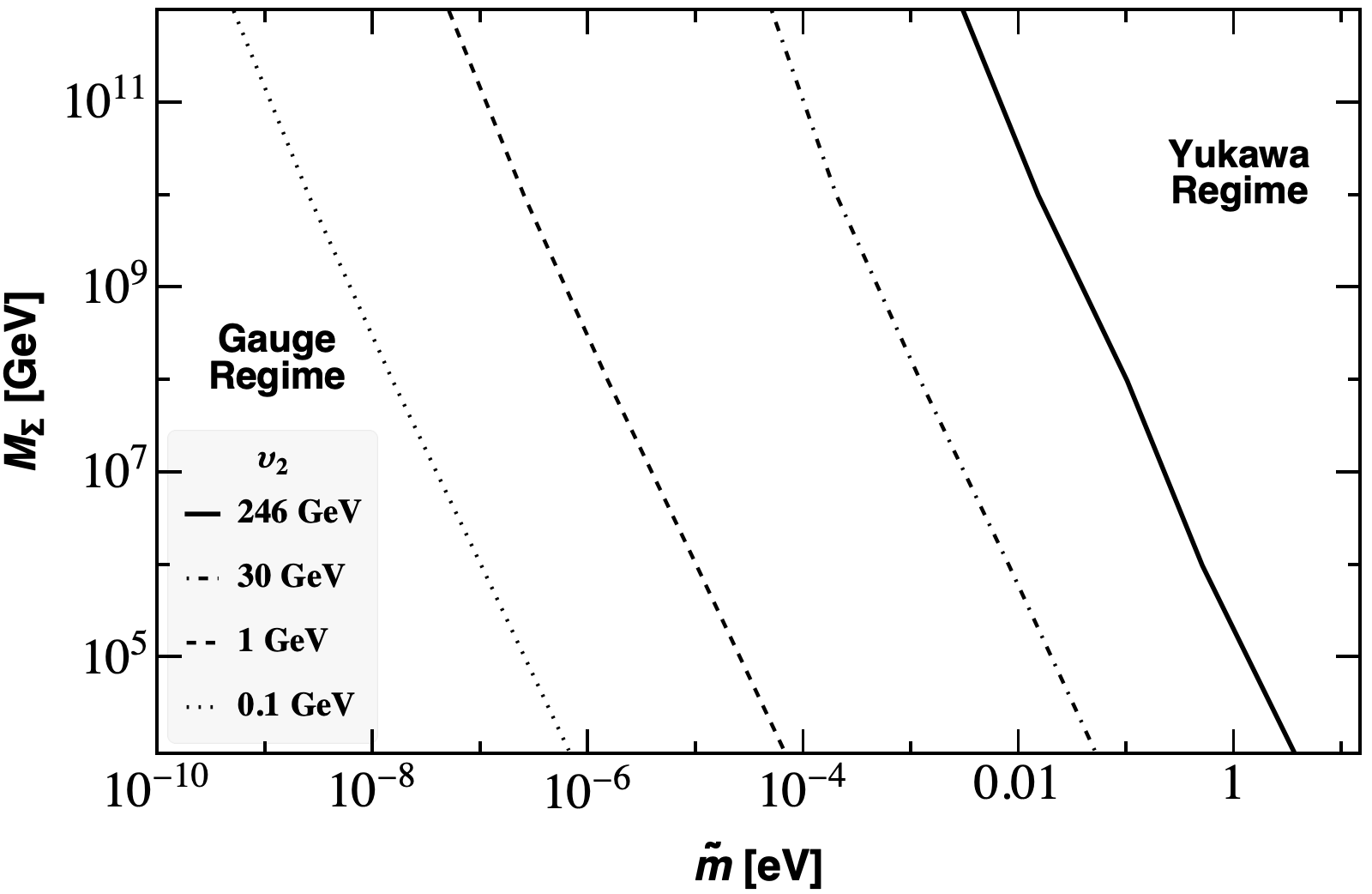}
\caption{In the lower (upper) $\mt$ region i.e. to the left (right) of each line corresponding to different VEVs, the gauge (Yukawa) interactions are dominant. Lowering the VEV shifts the line delimiting the two regimes towards smaller $\mt$ values. Hence, with smaller VEVs, unless $\mt$ is very small, one always remains in the Yukawa regime.}
\label{fig:gvsyu}
\end{figure} 

Therefore, we can define a minimum value of $\mt \equiv \mt_A$~\cite{Hambye:2012fh} for a given value of $M_\S$, below (above) which the gauge interactions are more dominant (suppressed) than the Yukawa interactions, allowing us to define the lower $\mt$ ``Gauge regime'' where gauge reactions are dominant, and the larger $\mt$ ``Yukawa regime'' where gauge reactions can be neglected; see Fig.~\ref{fig:gvsyu}. In the former, efficiency has a strong dependence on $M_\S$ and in the latter, efficiency depends entirely on $\mt$, just like type-I leptogenesis. Maximum efficiency is seen at the juncture of the two regimes. $\mt_A$ can be computed by requiring 
\begin{equation}
\frac{\g_D}{n_l^{eq} H}\bigg|_{z=z_A}=1\,.
\end{equation}

\section{Results and discussion}
\label{sec:result}

We now discuss the implications for triplet leptogenesis in presence of another Higgs doublet acquiring a tiny VEV. Note that, for smaller VEVs, even smaller values of $\mt$ give a higher decay or inverse decay rate than that of the standard VEV with higher $\mt$; see the purple lines corresponding to $\ut =1 {\rm~GeV}$ and $\mt = 0.0005 {\rm~ev}$ in Fig.~\ref{fig:int}. Hence, as one decreases the VEV, the Yukawa interactions start dominating over gauge interaction at smaller values of $z$ for a given value of $\mt$ and $M_\S$. While lower values of $z_A$ imply there's more triplet population left to produce the required asymmetry, the inverse decay rate is also enhanced and they can stay in equilibrium for $z>z_A$, suppressing the asymmetry generation. A direct consequence of lowering the VEV is the shifting of $\mt$ scale separating the two regimes, since the Yukawa interactions are inversely proportional to $\ut^2$, as can be seen in Fig.~\ref{fig:gvsyu}.

\subsection{Lower bound on triplet mass}
\label{sec:lowerbound}

\begin{figure*}[!htb]
\centering
\includegraphics[scale=0.4]{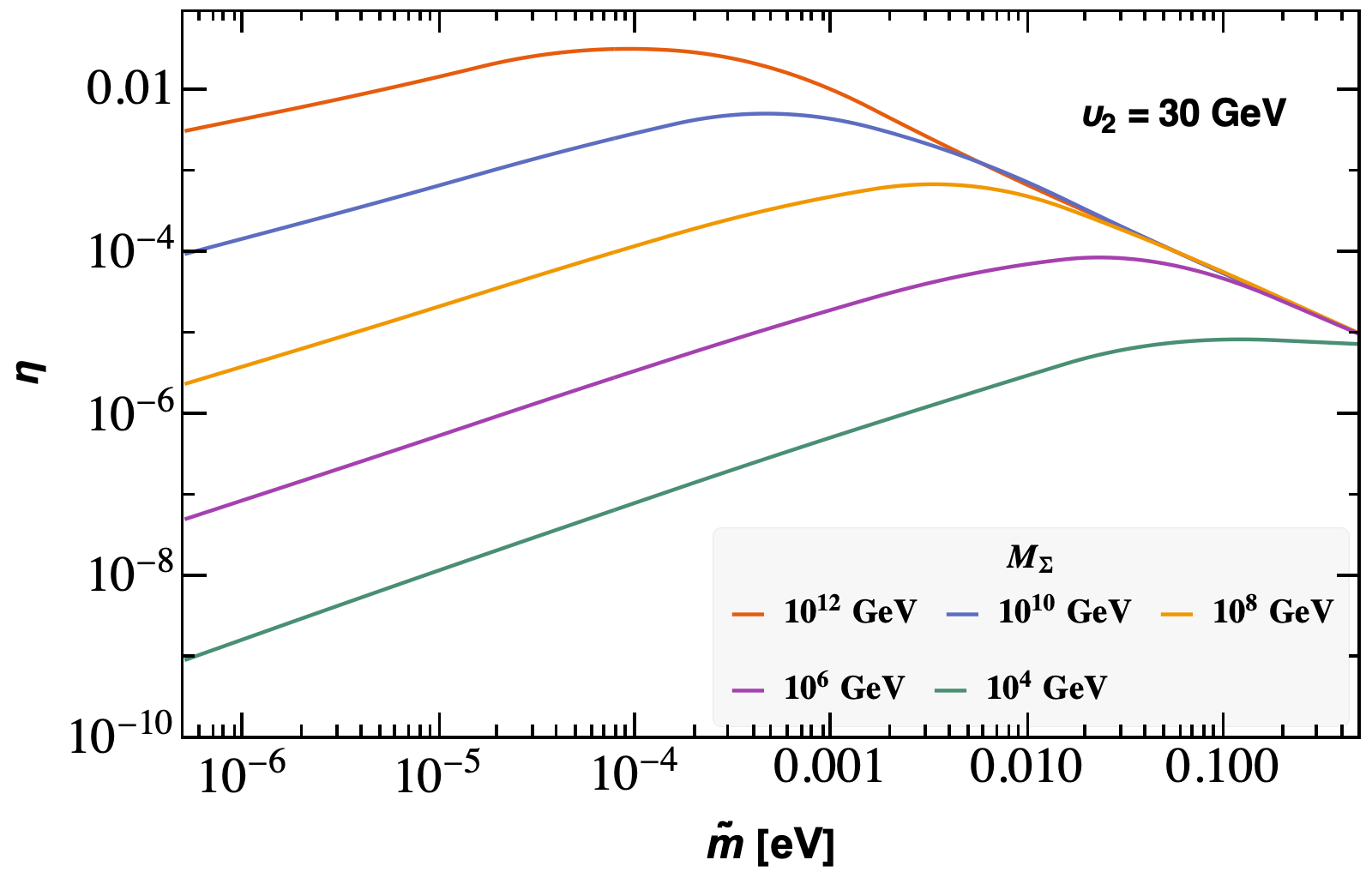}
\includegraphics[scale=0.4]{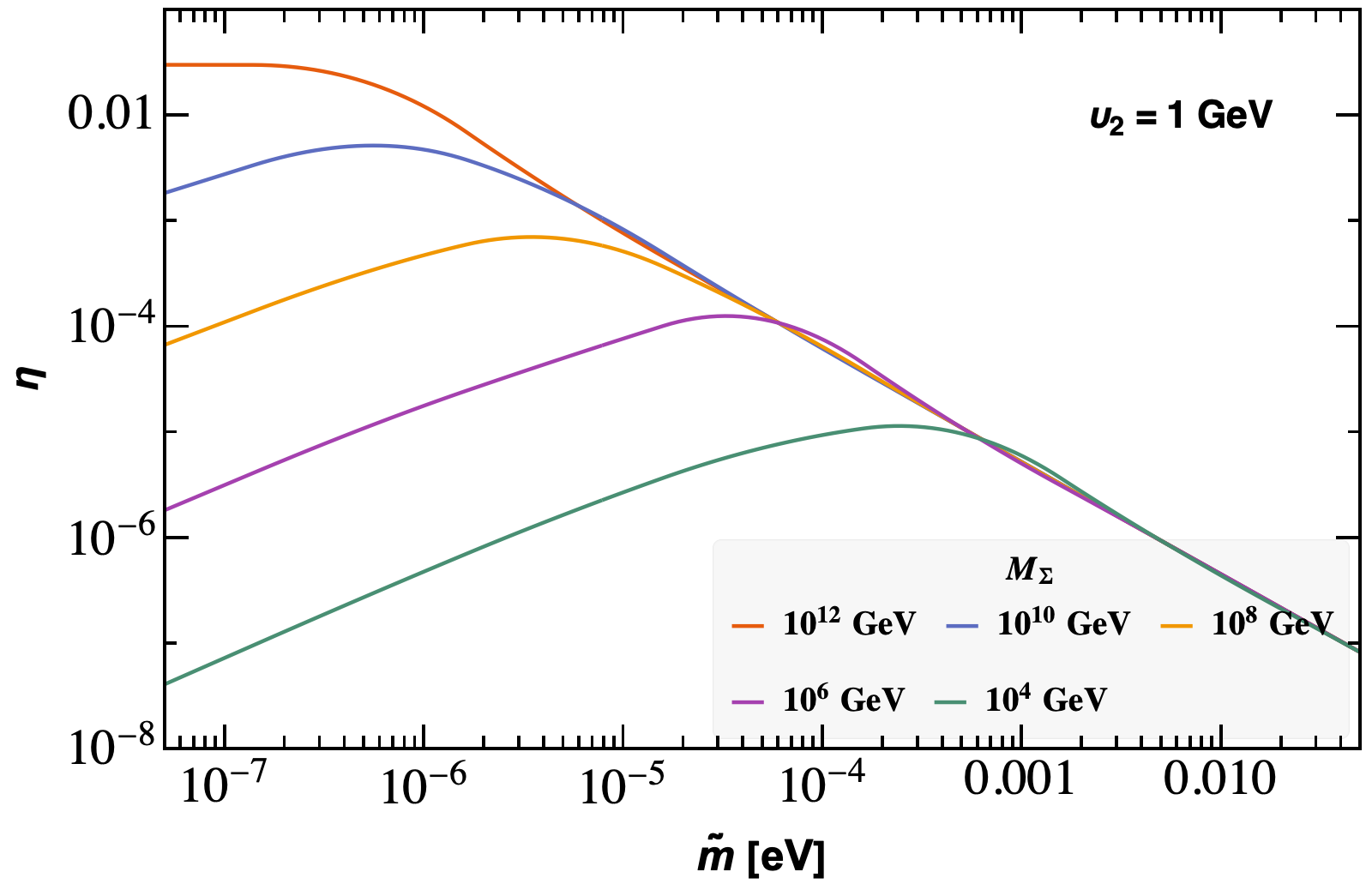}
\caption{$\eta$ as a function of $\tilde{m}$ and $M_\S$ for different values of $\ut$.}
\label{fig:asymm}
\end{figure*}

As discussed above, in order put a lower bound on $M_\S$, we need to determine the efficiency $\eta$ by numerically solving the Boltzmann equations, Eqs.~(\ref{eq:beqs}) and (\ref{eq:beql}). Considering only gauge scatterings and $\mathcal{O}(Y^2)$ interactions, the efficiency factor corresponding to different values of $\mt,M_\S$ and $\ut$ is shown in Fig.~\ref{fig:asymm}. The efficiency reaches a maximum at $\mt = \mt_A$ for different values of $M_\S$. It can be seen that for higher values of $\mt$, efficiency is independent of $M_\S$. We also find that lowering the VEV does not affect the maximum value of efficiency obtained for a particular $M_\S$ significantly; however, maximum efficiency is obtained for lower values of $\mt$ for smaller values of $\ut$. 

\begin{figure}[!htb]
\centering
\includegraphics[scale=0.4]{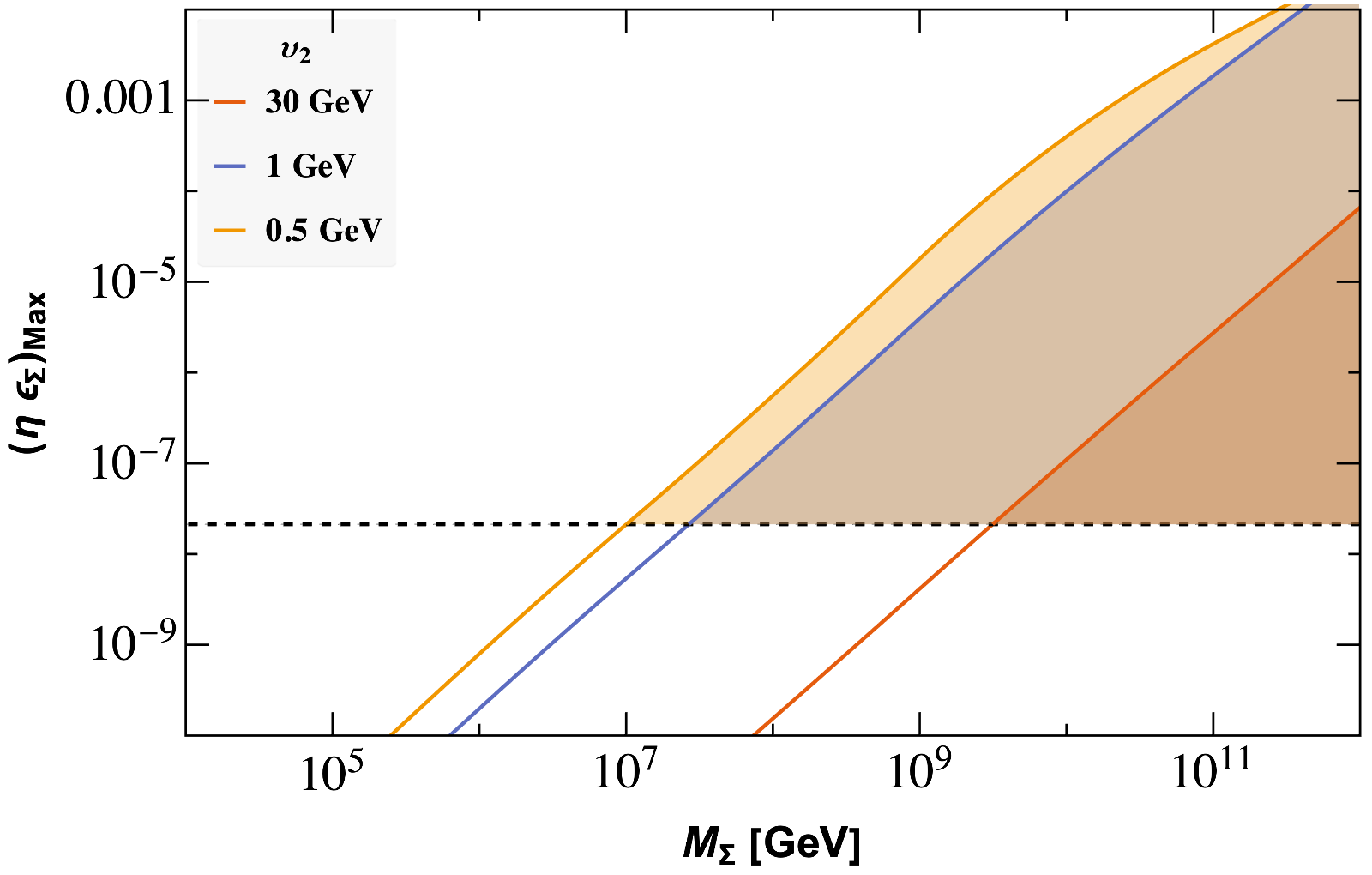}
\caption{The maximum value of $\eta\,\ve_\S$ one can obtain for different values of $M_\S$ and $\ut$. The dashed black line indicates the value of $\eta\, \ve_\S=(2.13 \pm 0.06) \times 10^{-8}$ required to produce the observed baryon asymmetry of the Universe in type-III leptogenesis with two Higgs doublets.}
\label{fig:parspace}
\end{figure}

In conjunction with the maximum bound on CP asymmetry, Eq.~(\ref{eq:DI}), which is relaxed due to lower VEVs, we plot the maximum of $\eta\, \ve_\S$ that one can obtain for a particular value of $M_\S$ and $\ut$ in Fig.~\ref{fig:parspace}. We also show the value of $\eta \ve_\S$ required to produce the observed baryon asymmetry from Eq.~(\ref{eq:yb}). 

Figure \ref{fig:parspace} is the main result of this work, allowing us to read the lower bound on $M_\S$ for successful leptogenesis,
\begin{align}\label{eq:bound}
M_\S > \begin{cases}
3 \times 10^{9} {\rm~GeV} {\rm~for~} \ut = 30 {\rm~GeV}\\
3 \times 10^{7} {\rm~GeV} {\rm~for~} \ut = 1 {\rm~GeV}\\
1 \times 10^{7} {\rm~GeV} {\rm~for~} \ut = 0.5 {\rm~GeV}
\end{cases}\,.
\end{align}

\subsection{Effect of $\Delta L=2$ scatterings}
\label{sec:l2scattering}

Now we consider the effect of $\Delta L =2$ scatterings mediated by right-handed neutrinos ($\mathcal{O}(Y^4)$), proportional to $M_\S^2 \bar{m}^2/\u^4$, where $\bar{m}^2=\sum m_i^2$ is the neutrino mass scale $\gtrsim 0.05 {\rm~eV}$. They contribute to the washout of lepton asymmetry and are significant in the strong washout regime. The thermally averaged $\Delta L=2$ scattering for $T \lesssim M/3$ in type-I leptogenesis has been approximated in Ref. \cite{Buchmuller:2004nz}. For type III, as interactions rates are multiplied by 3, we have
\begin{align}\label{eq:dell2}
\frac{\Gamma_{\Delta L=2}}{H} \approx \frac{3T}{2.2 \times 10^{13}{\rm~GeV}}\,\bigg(\frac{246 {\rm~GeV}}{\ut}\bigg)^4 \, \bigg(\frac{\bar{m}}{0.05 {\rm~eV}}\bigg)^2\,.
\end{align}
At $T \lesssim M_\S/3$, if these scatterings are in equilibrium, they can potentially wash out the asymmetry; however, this depends on the value of $\mt$. With hierarchical neutrino mass spectrum and $\mt \lesssim 10^{-1}{\rm~eV}$, these scatterings do not affect the final results unless $M_\S > 10^{14}{\rm~GeV}$. For $\mt \sim 1 {\rm~eV}$, these scatterings affect the results for $M_\S > 10^{12}{\rm~GeV}$.

In our model with smaller VEVs, these scatterings become more important than the standard case, as they are inversely proportional to $\ut^4$. The corresponding $\mt$ scale for a particular value of $M_\S$ above which these scattering can potentially wash out the generated asymmetry shifts by $(\ut/\u)^4$. Since, we observe that the maximum efficiency for a given triplet mass is seen toward low $\mt$ values where these scatterings are not very significant, the bounds in Eq.~(\ref{eq:bound}) remain valid. Though, in order to avoid too strong washout, these scatterings must go out of equilibrium at the sphaleron decoupling temperature around $T \sim 100{\rm~GeV}$, which leads to a lower bound on the VEV of $\pt$: $\ut \gtrsim 0.47$ from Eq.~(\ref{eq:dell2}).

\subsection{Effect of flavor}
\label{sec:flavor}

The flavor-blind analysis is justified for decaying right-handed neutrino masses above $10^{12}{\rm~GeV}$ as the flavor effects become important at temperatures below this value~\cite{Barbieri:1999ma, Abada:2006ea}. In the window $10^{9} - 10^{12}{\rm~GeV}$, the $\tau$ Yukawa interactions come into thermal equilibrium and the $B-L$ asymmetry is distributed into $L_\tau$ and $L_1$ ($L_e+L_\mu$). Consequently, the Boltzmann equation for $B-L$ asymmetry, Eq.~(\ref{eq:beql}),  splits into $Y_{L_\tau}$ and $Y_{L_1}$, making it a two-flavor problem. For temperatures below $10^{9}{\rm~GeV}$, the $\mu$ Yukawa interactions also come into thermal equilibrium and one has to consider three independent Boltzmann equations for $Y_{L_{e,\mu,\tau}}$. The flavor-dependent Boltzmann equation for lepton asymmetry can be written as~\cite{AristizabalSierra:2010mv}
\begin{equation}
\frac{dY_{\D_\a}}{dz}=-\frac{\gamma_D}{sHz} \bigg[\ve_\S^{l_\a}\bigg(\frac{Y_\S}{Y_\S^{eq}}-1\bigg)+\frac{K_\a}{2Y_l^{eq}}\small{\sum_{\b=e,\mu,\tau}}C_{\a\b}^l Y_{\D_\b} \bigg]\,,
\end{equation} 
where $\D_\a=Y_{B/3-L_\a}$ and $Y_{L_\a}=2Y_{l_\a}+Y_{e_\a}$. The flavor projector $K_\a$ is defined as
\begin{equation}
K_\a=\frac{Y_{\a 1}^\ast Y_{\a 1}}{(Y^\dagger Y)_{11}}=\frac{\mt_\a}{\mt}\,,
\end{equation}
and $\sum_{\a}K_\a=1$. The coefficients $C_{\a\b}^l$ couple the differential equations for different flavors and are determined from the reactions that are in equilibrium for a given temperature regime. The total asymmetry is then given by $Y_{B-L}=\sum_\a Y_{\D_\a}$.

Depending on the regime in which asymmetry is generated, the flavor effects allow us to obtain a less suppressed efficiency. Since the suppression in the gauge regime comes from flavor-blind gauge scatterings, flavor effects play a negligible role there. However, they can become important and sizable in the Yukawa regime where inverse decays are important. For example, if
\begin{equation}\label{eq:flavcon}
\ve_\S^{l_\b}\gg \ve_\S^{l_\a} {\rm~and~} \Gamma(\S \r {L_\a} H) \gg \Gamma(\S \r {L_\b} H)
\end{equation}  
i.e., the decay of the lightest triplet has a larger CP asymmetry and a smaller decay width in a particular flavor channel, then the suppression from inverse decays in that channel becomes negligible and we obtain a larger asymmetry than the unflavored case. 

\begin{figure}[!htb]
\centering
\includegraphics[scale=0.4]{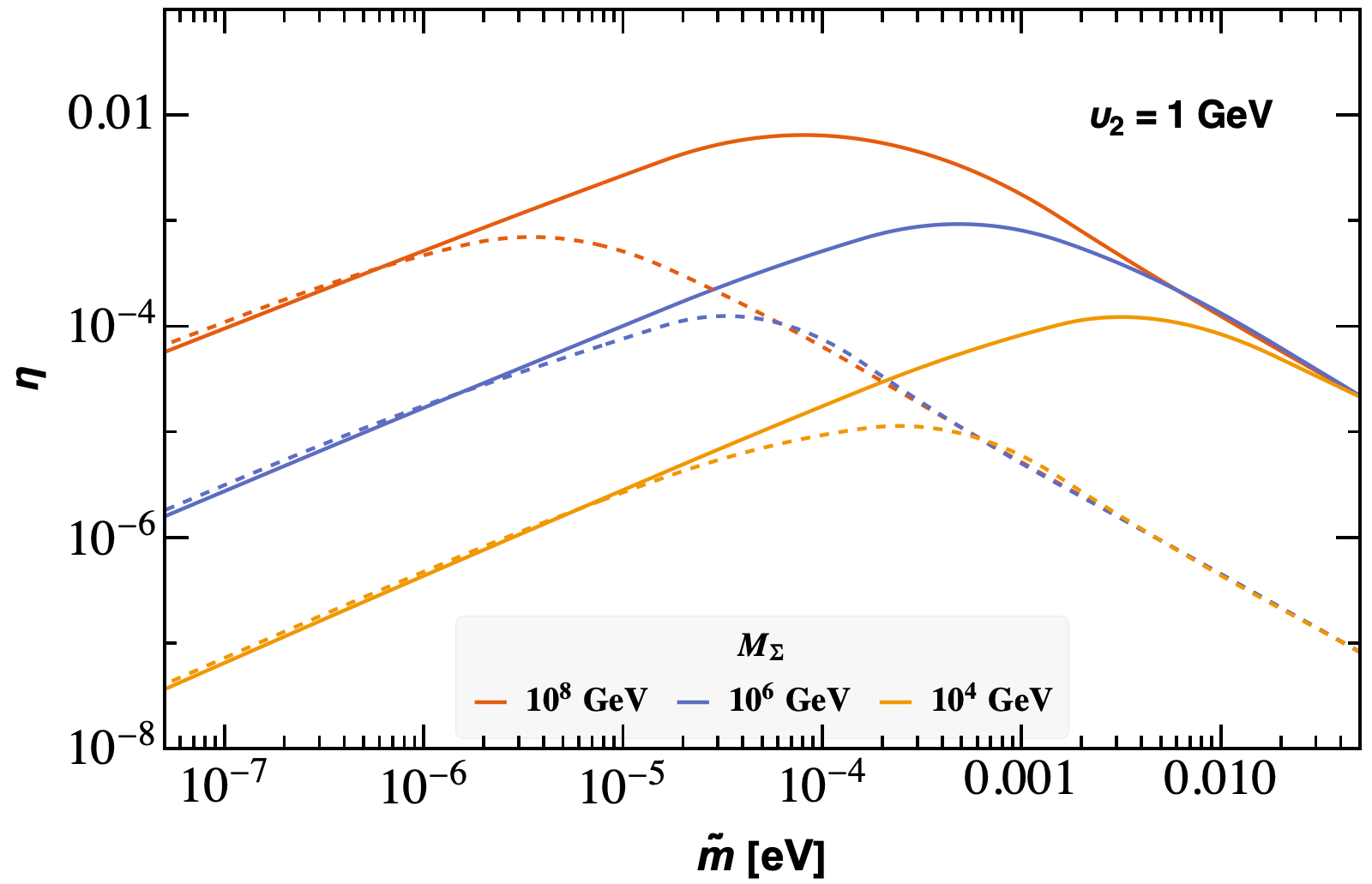}
\caption{$\eta$ as a function of $\mt$ for the flavor-dependent (solid curves) and flavor-independent (dashed curves) cases  assuming $K_1=0.99$, $K_\tau=0.01$ and $\ve_\S^{l_\tau}=1.1 \times \ve_\S$, $\ve_\S^{l_1}=-0.1 \times \ve_\S$.}
\label{fig:flavor}
\end{figure}

For simplicity and to illustrate the effect of flavor in both the regimes, we first work in the two-flavor framework ($L_\tau, L_1$) and follow closely the work in Ref.~\cite{AristizabalSierra:2010mv}. The flavor coupling matrix is given by~\cite{Nardi:2006fx}
\begin{equation}
C^l=\frac{1}{460}\bmt
196 & -24 \\
-9 & 156
\emt \,.
\end{equation}
Assuming $K_1=0.99$, $K_\tau=0.01$ and $\ve_\S^{l_\tau}=1.1 \times \ve_\S$, $\ve_\S^{l_1}=-0.1 \times \ve_\S$, we show the results for three different values of $M_\S$, $\ut=1 {\rm~GeV}$ in Fig. \ref{fig:flavor}. 

For a given value of $M_\S$ and $\mt$, it can be seen that in the gauge regime, the efficiency is roughly of the same order, whereas, for larger values of $\mt$, the efficiency can be enhanced by roughly 1-2 orders of magnitude. In conjunction with the bound on CP asymmetry, this allows one to reduce the bound on the triplet mass for successful leptogenesis in Eq. (\ref{eq:bound}) by an order of magnitude. Therefore, for a hierarchical spectrum of  decaying fermion triplets, we can lower the scale to roughly $\mathcal{O} (10^6 {\rm~GeV})$ for $\ut \sim 1 {\rm~GeV}$ once the flavor effects are included.

We have also checked the results for lower temperature regimes, where the three-flavor decomposition becomes important, and found that efficiency enhancement is of the same order as that of the two-flavor case; hence, the bound $M_\S > 10^6 {\rm~GeV}$ remains valid for lower temperatures as well, provided that Eq. (\ref{eq:flavcon}) is satisfied.

\subsection{Naturalness}
\label{sec:naturalness}

Having discussed the bound on triplet mass for successful leptogenesis, here we focus our attention toward naturalness, similar to the studies done for type-I and type-III seesaws in the literature \cite{Clarke:2015gwa,Clarke:2015hta,Goswami:2018jar}, for completeness. The high scale of the seesaw mediators and the presence of heavy particles in the model can provide large corrections to the Higgs mass, thus inducing the naturalness problem \cite{Vissani:1997ys}. Naturalness demands that these corrections to the Higgs mass are less than $\mathcal{O}({\rm TeV}^2)$ and leads to a conservative naturalness bound on the masses of the heavy particles.

\begin{figure}[!htb]
\centering
\includegraphics[scale=0.4]{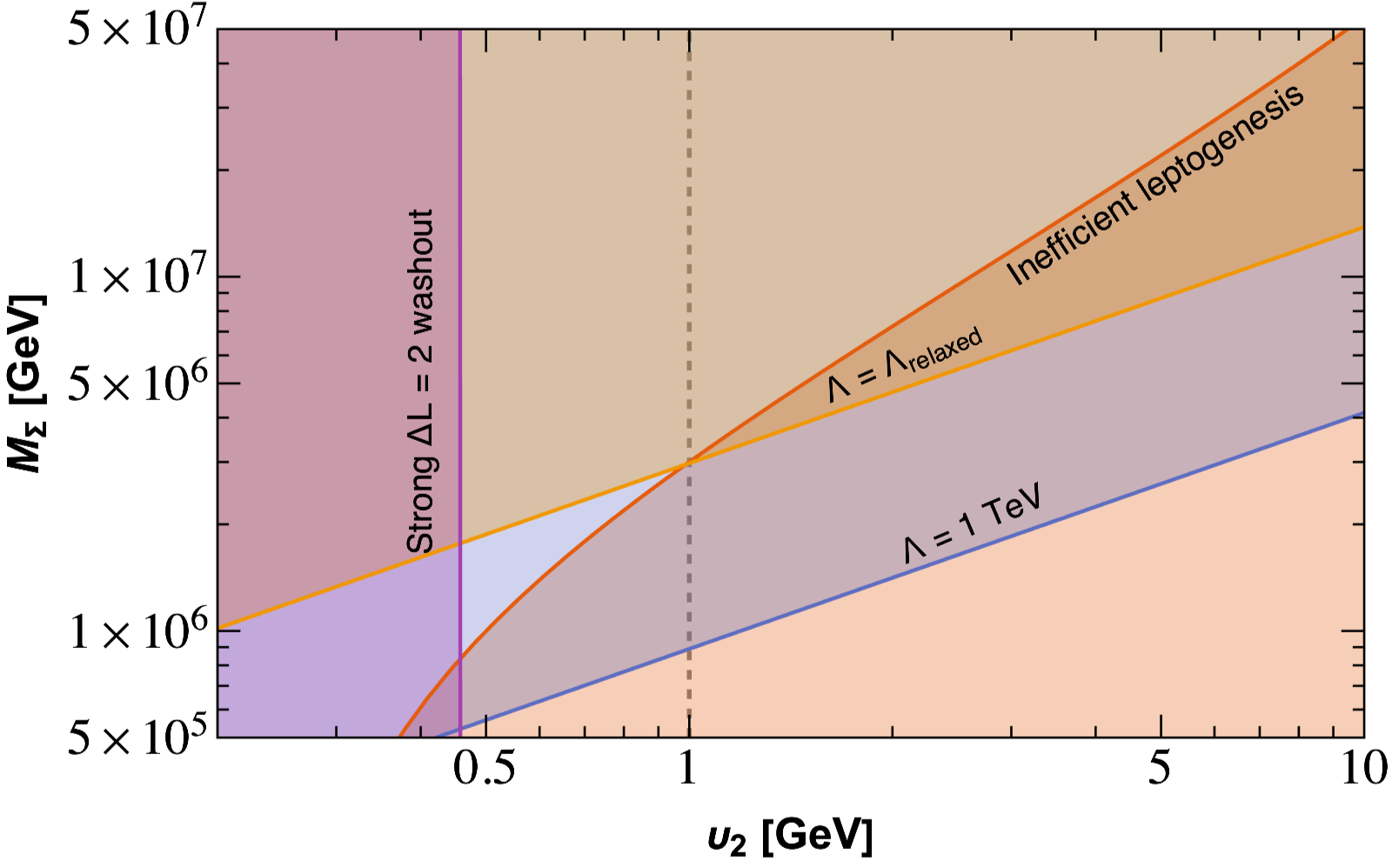}
\caption{Bounds on the triplet mass and the VEV of the second Higgs doublet from naturalness (blue and orange curves) and leptogenesis (red and purple curves). The gray dashed line indicates the value of $\ut$ below which the leptogeneis and relaxed naturalness bounds become compatible. }
\label{fig:naturalness}
\end{figure}

The naturalness constraints in the present setup involve the influence of the mass of the second Higgs doublet ($m_{22}$) on the first ($m_{11}$) and the influence of triplet masses ($M_\S$) on that of the second Higgs doublet ($m_{22}$). For the former, the authors of Ref.~\cite{Clarke:2015hta} find that a TeV-scale $m_{22}$ is completely natural. For the latter, the corrections to $m_{22}^2$ from $M_\S$ and the naturalness bound can be computed as
\begin{equation}
\delta m_{22}^2 \approx \frac{3}{4\pi^2}\,{\rm Tr}[{\bf Y}^\dagger \mathcal{D}_M^2 {\bf Y}] < \Lambda^2\,,
\end{equation}
where $\Lambda$ denotes the scale of corrections and $\mathcal{D}_M = {\rm diag}(M_{\S_1},M_{\S_2},M_{\S_3})$. Similar to the correction to the Higgs mass, if we take $\Lambda = 1$ TeV, the bound on $M_{\S}$ becomes \cite{Clarke:2015gwa}
\begin{equation}
M_\S \lesssim 3.5 \times 10^7 {\rm~GeV} \left(\frac{\ut}{246}\right)^{2/3}\,.
\end{equation}
However, since the heavy scalars are expected to be at the TeV scale, one can consider slightly relaxed bounds by taking the correction to $m_{22}^2$ larger than $\mathcal{O}({\rm TeV}^2)$. Following the approach in Ref.~\cite{Clarke:2015hta}, we take this relaxed bound $\Lambda_{\rm relaxed} = {\rm min}(10 {\rm~TeV}, 10\sqrt{m_h^2 \tan \beta /2})$, where we have used the consistency condition of Eq.~\eqref{eq:cons}. In Fig.~\ref{fig:naturalness}, we show these bounds in conjunction with the bounds from efficient leptogenesis including the flavor effects (a similar figure is provided for the type-I seesaw in Ref.~\cite{Clarke:2015hta}).

It can be seen that demanding corrections to $m_{22}$ of the order TeV is not consistent with the bounds from leptogenesis despite the lowering of the leptogenesis scale via the small VEV of the second Higgs doublet. However, if we consider the relaxed naturalness bound, a small region of parameter space opens up that is compatible with all the bounds (as indicated by the light blue region in the center) and can be read as $8.5 \times 10^5 {\rm~GeV} < M_\S < 3 \times 10^6 {\rm~GeV}$. Furthermore, in order for these bounds to be compatible, $\ut$ should be less than 1 GeV (indicated by the dashed gray line).

\section{Summary}
\label{sec:con}

We considered the possibility of lowering the scale of type-III leptogenesis in the context of a two-Higgs-doublet model. Because of the gauge scatterings of fermion triplets, the asymmetry produced by their decays is highly suppressed and successful leptogenesis is possible only for $M_\S > 3 \times 10^{10} {\rm~GeV} ~(10^9 {\rm~GeV})$ in single-flavor (three-flavor) approximation \cite{Hambye:2003rt}. We have shown that if all the heavy fermion triplets couple only to the second Higgs doublet, which obtains a small VEV, the absolute bound on the mass of the decaying triplet can be lowered to $~10^7 {\rm~GeV}$, roughly 3 orders of magnitude below the current bounds for successful leptogenesis. We also include the flavor effects and find that there can be a further reduction to $10^{6}$ GeV.

We have studied the interplay of gauge and Yukawa scatterings for different VEVs, and find the latter get more dominant than the former as one decreases the VEV. This results in shifting the scale delimiting the gauge and Yukawa regimes, at the junction of which one obtains maximum efficiency. Hence, the efficiency is enhanced for smaller values of $\mt$ compared to the standard case with just the SM Higgs doublet. This suggests that, for lower VEVs, the effective contribution to the lightest active neutrino needs to be smaller. Higher values of $\mt$ lead to significant suppression from inverse decays and the dynamics become similar to the case of type-I leptogenesis. The lowering of  the scale is mostly due to the relaxation of Davidson-Ibarra bound on the maximum CP asymmetry for smaller VEVs. Furthermore, constraints from strong washout due to $\Delta L =2$ scatterings and compatibility with naturalness impose stringent constraints on the VEV of the second Higgs doublet to be $0.47 {\rm~GeV} \lesssim \ut \lesssim 1 {\rm~GeV}$. 

The analysis above highlights that unlike the case of type-I leptogenesis where the scale can be lowered to TeV scales (and below) in the presence of a second Higgs doublet, there exists only a small region of parameter space around $M_\S \sim 10^6$ GeV that is compatible with both leptogenesis and naturalness.

\textbf{Acknowledgments} S.G. acknowledges the J.C. Bose Fellowship (JCB/2020/000011) of Science and Engineering Research Board of Department of Science and Technology, Government of India. D.V. is supported by the ``Generalitat Valenciana” through the GenT Excellence Program (CIDEGENT/2020/020) and greatly acknowledges the hospitality at Physical Research Laboratory and support via the J.C. Bose Fellowship (JCB/2020/000011) during the initial stage of the work. 

\bibliographystyle{utphys}
\bibliography{refs}

\end{document}